\documentstyle[12pt]{article}
\begin{document}
\begin{center}
 \huge Spacetime Defects: Torsion Loops
\end{center}
\vspace{3ex}
\centerline{\large \em Patricio S. Letelier\footnote{e-mail:
letelier@ime.unicamp.br} }
\vspace{1ex}
\begin{center}
 Departamento de Matem\'atica Aplicada-IMECC\\
Universidade Estadual de Campinas\\
13081-970 Campinas. S.P., Brazil \\ \vspace*{0.3cm} { \small April 19, 1995}
\end{center}

\baselineskip 0.7cm
Spacetimes with everywhere vanishing curvature tensor,  but with
torsion different from zero only on  world sheets that represent
closed loops  in ordinary space are presented, also defects along open
curves with end points at infinity are studied. The case of defects along
timelike loops  is also considered and the geodesics in these
 spaces are briefly discussed.

\vspace*{6ex}
 \newpage
\baselineskip 0.9cm

Conical singularities or spacetime defects are  characterized
by Riemann-Christoffel curvature tensor, or Cartan torsion,  or
 both different from zero only on the subspace (event, world line, world
sheet, or world tube) that describes the evolution of the defect (texture,
monopole, string, or membrane). In other words we have
that the curvature, the torsion, or both, are proportional
 to distributions with support on the defect.
Spacetimes with conical  singularities
of different types has been studied recently in a variety of contexts, e.g.
spinning  strings with cosmic dislocations \cite{gallet}\cite{lchi},
pure spacetime dislocations \cite{tod} and \cite{edelen}, also in low
 dimensional gravity \cite{koh}. For the discussion of a great variety
of defects see Ref. \cite{lew}.

 The already known cases of line defects, in  usual  space, are
 infinite strait line defects with or without torsion. The case of
 constant torsion was studied in \cite{gallet}.  Lines  of
 dislocations  that grow  linearly with time  and also spin
 that grows linearly along the defect were studied in  \cite{tod} and
 \cite{edelen}. It is worth to mention that following the same methodology of
 \cite{lchi}  one can also super impose
 to this defect   the defect angle
 that  represents a disclination (cosmic string). The
 purpose of this letter is to study spacetimes  with torsion line
 defects of arbitrary shape.

Let us consider the metric
\begin{equation}
ds^2=(\omega^0)^2 -(\omega^1)^2-(\omega^2)^2-(\omega^3)^2 , \label{m0}
\end{equation}
with
\begin{eqnarray}
\omega^0&=&dt +B_x dx +B_y dy +B_z dz ,\nonumber\\
        & =&dt +{\bf B}\cdot d {\bf x} \nonumber\\
\omega^1&=&dx, \,\,\, \omega^2=dy, \,\,\, \omega^3 = d z \label{f1}
\end{eqnarray}
where ${\bf B}=(B_x,B_y,B_z)$ are arbitrary
functions of the coordinates ${\bf x}= (x,y,z)$ only. Let us assume
that the spacetime has a torsion of the form,
\begin{eqnarray}
T^0&=&J_z \omega^1\wedge\omega^2 +J_x \omega^2\wedge\omega^3
+J_y\omega^3\wedge\omega^1, \nonumber\\
T^1&=&T^2 =\, T^3 =0 , \label{t1}
\end{eqnarray}
where ${\bf J}=(J_x,J_y,J_z)$ are functions of ${\bf x}= (x,y,z)$ only. From
the first Cartan structure equation,
\begin{equation}
\b T^{a}=d \omega^{a} +\omega^{a}_{\,\,\,b}\wedge
 \omega^{b}, \label{1car}
\end{equation}
we find the connection one-forms,
\begin{eqnarray}
2\omega^{01}&=&(\partial_x B_y -\partial_y B_x -J_z)\omega^2 +
(\partial_z B_x -\partial_x B_z -J_y)\omega^3 , \nonumber\\
2\omega^{02}&=&-(\partial_x B_y -\partial_y B_x -J_z)\omega^1 +
(\partial_y B_z -\partial_z B_y -J_x)\omega^3 , \nonumber\\
2\omega^{12}&=&(\partial_x B_y -\partial_y B_x -J_z)\omega^0 ,\nonumber\\
2\omega^{23}&=&(\partial_y B_z -\partial_z B_y -J_x)\omega^0 ,\nonumber\\
2\omega^{31}&=&(\partial_z B_x -\partial_x B_z -J_y)\omega^0 .\label{con1}
\end{eqnarray}
Now, we shall postulate that the ``Cartesian" vectors ${\bf B}$ and
${\bf J}$ are related by the Cartesian equation,
\begin{equation}
\nabla \times {\bf B} = {\bf J}. \label{bio}
\end{equation}
Then, we have that the connection one-forms are zero, $\omega^{ a b}=0$,
and as a consequence of the second Cartan structure equations,
\begin{equation}
R^{a}_{\,\,\,b} =d\omega^{a}_{\,\,\, b} +
 \omega^{a}_{\,\,\,c}\wedge
 \omega^{c}_{\,\,\, b}, \label{car2}
\end{equation}
the curvature two-forms are also zero, $R^{ab}=0$. Thus we have a
 spacetime with zero curvature and non vanishing
 torsion, i.e., a Weitzenb\"ock
space \cite{schou}. We can choose the torsion (\ref{t1}) as a
distribution with support  along a curve $C$ with parametric equation
${\bf x}'={\bf x}'(\lambda)$ taking the function ${\bf J}$ as
\begin{equation}
{\bf J}({\bf x})= I \int_C \delta^{(3)}({\bf x} -
{\bf x}'(\lambda))\frac{d{\bf x}'(\lambda)}{d \lambda} d \lambda ,
\label{j1}
\end{equation}
where $I $ is an arbitrary constant.  The components of ${\bf J}$ can also
be written in the more appealing form \cite{klein},
\begin{equation}
J_k({\bf x})=I\delta^{(2)}_{k}({\bf x}_\perp, C).
\label{jdel}
\end{equation}
The function (\ref{j1}) isline $C$ that generalizes to an arbitrary curve the
formula valid for
a strait line of torsion placed along the $z$-axis
\begin{equation}
J_z=I\int\delta(x)\delta(y)\delta(z-z')dz' =I\delta(x)\delta(y).\label{j2}
\end{equation}
The integrability condition for  equation (\ref{bio})
is $\nabla\cdot {\bf J}=0$; from
(\ref{j1}) we get,
\begin{equation}
\nabla\cdot{\bf J} = \delta^{(3)}({\bf x} -{\bf x}'_i)
 -\delta^{(3)}({\bf x} -{\bf x}'_f),
 \label{if}
\end{equation}
Thus to satisfy this  condition we need  either that the initial and final
points of the curve coincide, ${\bf x}'_i ={\bf x}'_f$,
i.e.,  a closed curve, or that the two end points of $C$ be at
infinity. In both cases we find that the solution of (\ref{bio}) is given
by the well known formula,
\begin{equation}
{\bf B}= \frac{I}{4\pi} \int_C \frac{d {\bf x}'\times({\bf x}-{\bf x}')}{ |
({\bf x}-{\bf x}'|^3}.
\label{B}
\end{equation}
  The metric (\ref{m0}) built with the forms (\ref{f1}) with
 the functions ${\bf B}$  given above have zero curvature tensor
 and non vanishing torsion along the curve $C$ of the usual space. In the four
dimensional space we have a cylindrical surface with generators parallel
to the $t$-axis that pierce the hyperplane $t=$constant in $C$.

Another spacetime  with the same characteristics is (\ref{m0})
with the one-forms
 \begin{eqnarray}
\omega^0&=&dt, \,\,\, \omega^1=dx, \,\,\, \omega^2 = dy \nonumber\\
\omega^3&=&dz +E_x dx +E_y dy +E_t dt ,\label{f2}
\end{eqnarray}
where ${\bf E}=(E_x,E_y,E_t)$ are arbitrary
functions of the coordinates  $(x,y,t)$, and  torsion,
\begin{eqnarray}
T^0&=&T^1=\,T^2=0, \nonumber\\
T^3&=&K_t \omega^1\wedge\omega^2 +K_x \omega^2\wedge\omega^0
+K_y\omega^0\wedge\omega^1, \label{t2}
\end{eqnarray}
where the functions   ${\bf K}=(K_x,K_y,K_t)$ are arbitrary
functions connection one-forms gives us,
 \begin{equation}
\tilde\nabla \times {\bf E} =
{\bf K},\,\,\,\, (\tilde\nabla=(\partial_x,\partial_y,\partial_t)).
\label{bio2}
\end{equation}
Therefore, solutions of this last equation for a  torsion built with a function
${\bf K}$  similar to (\ref{j1}) but
 with the coordinate $z$ changed by $t$ are spacetimes
with everywhere vanishing curvature tensor, and torsion given by
 a distribution with support on a curve $\tilde C$ in the three
space $(x,y,t)$.  In the four dimensional spacetime this curve are
lifted to a cylindrical surface with generators parallel to the $z$-axis.

Now we want to consider test particles moving in these spacetimes
 with torsion defects.
We have two different equations of motion: the autoparallels and the
 geodesics. Since,
in  both classes of spacetimes   we have zero Riemann-Cartan
 connection,  autoparallels are strait lines. For the first type of defects we
have
 $\nabla\times{\bf B}=0$ outside the defect. Then the geodesic equation, in
 this case, reduces to
\begin{equation}
\dot{t}+{\bf B}\cdot \dot{\bf x}=W, \,\,\, \ddot{\bf x}=0, \label{geo1}
\end{equation}
where the dots denote differentiation  with respect the proper time $s$; $W$ is
 an integration constant. Hence, a particle following a geodesic travels
 along a strait
line, in general,  with variable speed. For the second case, outside the
 defect we also have $\tilde\nabla\times{\bf E}=0$ and the geodesic
 equation gives us,
\begin{eqnarray}
&&\ddot{t}=\ddot{x}=\ddot{y}=0,\nonumber\\
&&\dot{z}=\tilde{W}-E_x\dot{x}-E_y\dot{y}-E_t\dot{t},\label{geo2}
\end{eqnarray}
where $\tilde{W}$ denotes another integration constant. In this case the
geodesics are  plane curves,  the $z$-axis is also contained on the
 plane defined by the curve.

In summary, we havby everywhere zero Riemann-Christoffel curvature tensor and
non vanishing
torsion.  The torsion can be concentrated on a closed line or on an open
line with end points at infinity. Ideas for  a field theoretical approach to
this defects were advanced in \cite{lchi}, we hope to comeback to this
point in another occasion.

\end{document}